# In Silico Fatigue Optimization of TAVR Stent Designs with Physiological Motion in a Beating Heart Model

Authors: Kyle Baylous[1], Ryan Helbock[1], Brandon Kovarovic[1], Salwa Anam[1], Marvin Slepian[2], Danny Bluestein[1]*

[1]Department of Biomedical Engineering, Stony Brook University, Stony Brook, NY, USA 11794

[2]Department of Medicine and Biomedical Engineering Sarver Heart Center, University of Arizona, Tucson, AZ 85721

\* Corresponding Author Contact Information

Dr. Danny Bluestein

Department of Biomedical Engineering, Stony Brook University

T8-050 Health Sciences Center

Stony Brook, NY 11794-8084, USA

Tel: +1 (631) 444-2156

Email: danny.bluestein@stonybrook.edu

**ABSTRACT**

The rapid expansion of TAVR to younger, low-risk patients raises concerns regarding device durability. Necessarily, extended stent lifetime will become more critical for new generation devices. *In vitro* methods commonly used for TAVR stent fatigue testing exclude the effects of the beating heart. We present a more realistic *in silico* stent fatigue analysis utilizing a beating heart model in which TAVR stents experience complex, nonuniform dynamic loading. Virtual TAVR deployments were simulated in the SIMULIA Living Heart Human Model of a beating heart using stent models of the self-expandable nitinol 26-mm CoreValve and Evolut R devices, and a 27-mm PolyV-2. Stent deformation was monitored over three cardiac cycles, and fatigue resistance was evaluated for the nitinol stents using finite element analysis via ABAQUS/Explicit. In all models, there were elements in which strains exceeded fatigue failure. The PolyV-2 stent had far fewer failing elements since its struts were optimized to reduce the strain in stent joints, achieving better fatigue resistance in the stent crown and waist elements. Different stent sections showed markedly different fatigue resistance due to the varying loading conditions. This study demonstrates the utility of advanced *in silico* analysis of devices deployed within a beating heart that mimics *in vivo* loading, offering a cost-effective alternative to human or animal trials and establishing a platform to assess the impact of device design on device durability. The limited fatigue life of TAVR stents indicated here highlights a clinical complication that may eventually develop as younger, lower-risk TAVR patients, age.

**INTRODUCTION**

The well-established, minimally-invasive transcatheter aortic valve replacement (TAVR) procedure is performed to treat aortic stenosis (AS), the narrowing of the aortic valve opening [1]. AS results from progressive calcific aortic valve disease (CAVD), effecting over nine million patients globally [2]. TAVR is an alternative to open-heart surgical aortic valve replacement (SAVR), rapidly expanding to younger, lower risk patients [3]. Patients with severe AS at low surgical risk display improved outcomes with TAVR over SAVR [4]. However, TAVR device durability in these patients is a major concern. This highlights the importance of long-term fatigue resistance of device leaflets and stent frames as they must remain safe and functional for longer periods of time, or else risk patients with repeat replacement surgeries [5, 6]. Self-expanding TAVR devices use nitinol stents. Fracture occurrences of some nitinol-based endovascular stents have been reported after one year [7]. Stent failure results from *in vivo* cyclic displacements, highlighting TAVR design optimization and testing as it relates to safety of nitinol devices [3, 8, 9].

Testing standards for TAVR devices, chiefly the ISO 5840-3 [10], were created by combining the hydrodynamic accelerated wear leaflet testing standards of surgical valves an stent frame fatigue analysis of stent and graft devices. Leaflets are evaluated with various pressure loading conditions and deployment configurations for a minimum of 200M cycles [7, 10-12]. Stent frames are required to withstand 400M cycles with the manufacturer subject to "identify and justify appropriate *in vivo* loading conditions" [10]. Previous research has focused on induced strain on the stent frame from the flexure of the diastolic leaflets [13], compliant tubes with cyclic pressure waveforms [14], or uniform radial displacements measured from *in vivo* observations [7]. However, TAVR devices experience complex, nonuniform dynamic loading. There is a paucity of longer-term data from clinical studies indicating stent fracture risks within patients, with the research focus on structural valve degeneration (SVD) of the leaflet tissues [15]. However, TAVR stent durability may now be assessed via the Living Heart Human Model, or LHHM (Fig.

1), serving as a fully coupled, electromechanical beating heart model with realistic *in vivo* loading conditions [16].

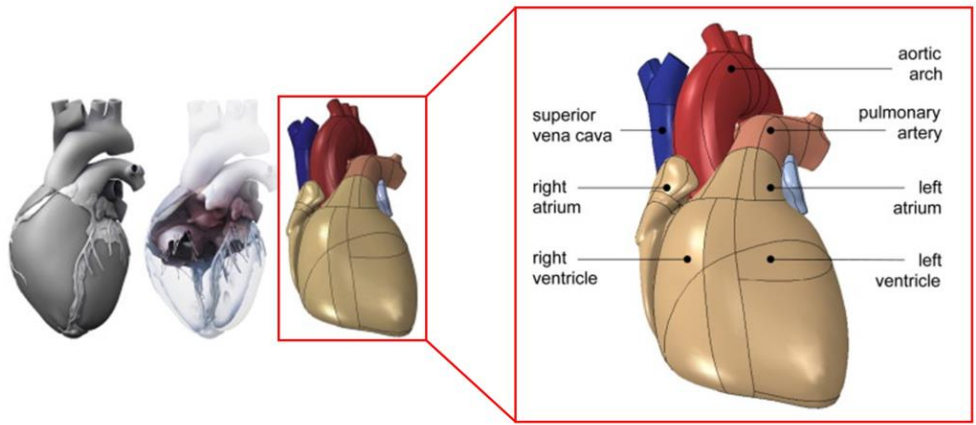

**Figure 1.** The electromechanically coupled Living Heart Human Model (LHHM) with various anatomic details including the ventricles, atria, aortic arch, superior vena cava, and pulmonary artery. Adapted from [16].

The LHHM has been used for analysis of ascending thoracic aortic aneurysm [17], examining left ventricular biomechanics associated with AS [18], patient-specific TAVR deployments [19, 20], and virtual coronary stent deployments to assess long-term durability [21]. Regarding TAVR stents, no such analysis has been conducted. Stent lifetime is becoming more critical with the rapid development of polymeric TAVR devices [22]. Therefore, in this study, TAVR stent fatigue is evaluated with a more accurate *in silico* approach by examining the interaction of multiple TAVR stent frames with the LHHM, thereby demonstrating the utility of advanced *in silico* analysis to overcome the limitations of current *in vitro* durability testing and address fatigue concerns.

**METHODS**

*TAVR Stent Models*

Stents models were generated for the commercial CoreValve and Evolut R self-expandable devices (Medtronic plc, Dublin, Ireland) after measuring struts of the 29-mm CoreValve via caliper and the acquisition of high-precision micro-CT images. Variable thickness was observed in the circumferential direction (0.2-0.37 mm) (Fig. 2a). All stents were assumed to have the same thickness dimensions due to the lack of published commercial data. The length of each strut was then calculated based off each stent's outer diameter and the number of cells using an in-house MATLAB code (MathWorks, Inc.). The repeating strut patterns of each stent model were dimensioned for device reconstruction via the computer aided design (CAD) software SOLIDWORKS (DS SolidWorks Corp., Waltham, MA). For validation, finite element analysis (FEA) of the radial force was conducted in ABAQUS/Standard (Dassault Systèmes, SIMULIA Corp.) and compared to *in vitro* results [23]. Superelastic nitinol material properties at 37°C were adjusted to achieve similar curves and mesh convergence study was performed (Fig. 2a) to optimize the simulations (~11K C3D8R elements per stent) [23].

The virtual stent models represented the commercial devices well after overlaying each on the rendered image of the physical stent (Fig. 3a). The virtual stents can be considered highly accurate replicas of their commercial counterparts when considering device geometry. Once the 26-mm CoreValve and Evolut R control models were generated, the strut widths were scaled up and down by 20% (Fig. 2b); three iterations for each stent – control, 20% increase and 20% decrease – were deployed virtually. Additionally, the novel 27-mm PolyV-2 (PolyNova Cardiovascular Inc, Stony Brook, NY) TAVR stent was analyzed using a virtual model provided by the company. No strut scaling was conducted for PolyV-2 as it was previously optimized to minimize the stresses and strains. The 26-mm CoreValve and Evolut R devices were chosen along with the 27-mm PolyV-2 for this analysis as the three devices are correctly sized to fit the aortic annulus dimensions of the LHHM.

*TAVR Stent Deployment*

The stents were deployed inside the LHHM v.2019 beating heart model (Dassault Systèmes, SIMULIA Corp., Johnston, RI) and simulated with ABAQUS/Explicit. Each stent was assigned identical superelastic nitinol material properties. The virtual TAVR deployments included crimping and positioning of each device, followed by removing the crimping sheath to deploy each stent in the native aortic valve of the LHHM (Supplementary Videos 1-3). Each stent was crimped to 16 French by applying a radial boundary condition on the nodes of the crimping sheath. An axial boundary condition was applied to the nodes of the stent and crimping sheath to position each device in the native aortic valve. The stents were then deployed by reversing the initial radial boundary condition on the nodes of the crimping sheath, allowing the stents to expand and deploy with an implantation depth between 3-6 mm. Contact was defined between the crimping sheath and the outer surface of each stent, as well as between the crimping sheath and the ventricular side of the aortic valve leaflets. Additionally, contact was defined between the stent, aortic arch and left ventricular outflow tract. After deployment, the heart was configured to beat three times. The deformation of the stent models was tracked over each of the three cycles until the deformations were repeatable [23]. As inclusion of the leaflets is computationally expensive and the effects on stent deployment have shown to be negligible [24] the prosthetic leaflets were excluded in order to focus on the TAVR stent frame fatigue. A variable mass scaling was used to achieve a stable time increment with minimal effects on the model dynamics [19, 23].

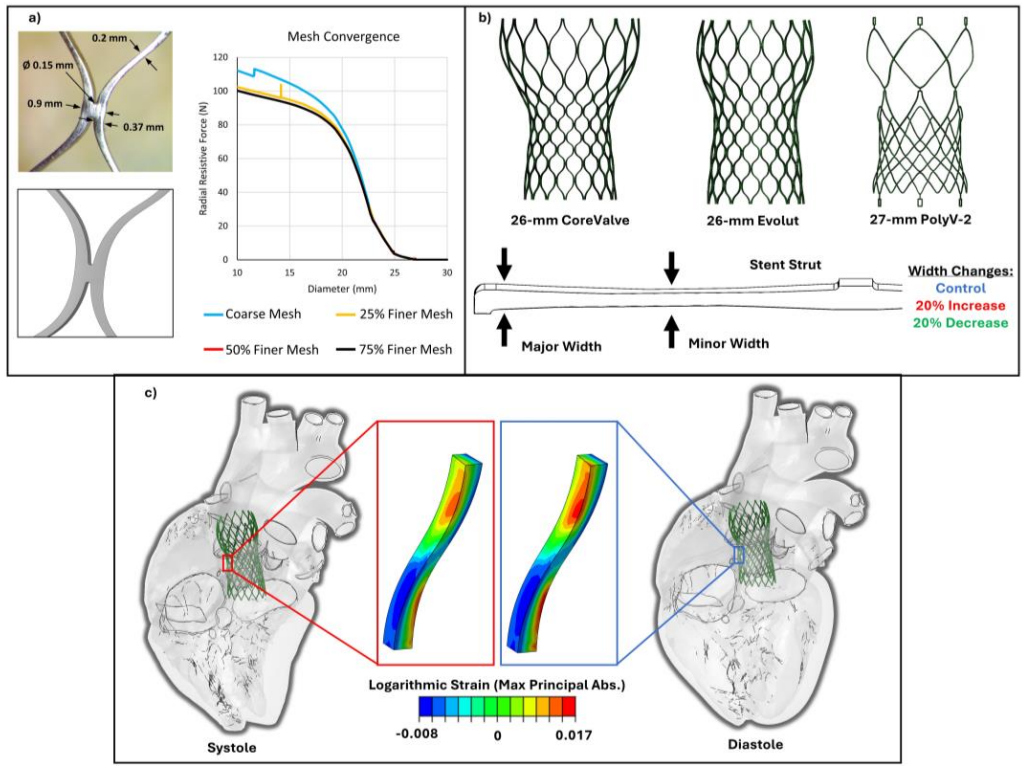

**Figure 2.** Workflow for TAVR stent fatigue analysis in the LHHM. a) Comparison of variable thickness in true stent model (29-mm CoreValve in image with relevant dimensions) and *in silico* stent model along with the mesh convergence study performed on a column of struts via radial force measurements. b) FE models of the 26-mm CoreValve, 26-mm Evolut and 27-mm PolyV-2 stent models: Control, 20% increase, and 20% decrease strut width versions created for each stent. c) *In silico* TAVR stent deployment for fatigue analysis.

*Fatigue Analysis of TAVR Stents*

FEA was utilized to determine fatigue resistance of the stents with an analysis for nitinol materials as described by Pelton et al. [7]. The principal logarithmic strains were extracted from each stent element (Fig. 2c) during the third beat cycle in ABAQUS. The mean strain and the strain amplitude of each element was computed and plotted on constant-life diagrams using MATLAB. Each data point on the fatigue plots represents the maximal mean strain and maximal strain amplitude - plotted along the x and y axes, respectively - recorded along the third cardiac cycle. Pelton et al. defined failure criteria – the "Pelton curve" - where absolute principal strains over 8% and strain amplitudes over 0.4% define failed stent elements. This failure threshold was overlaid on each plot to analyze "failed" stent elements for each model, exceeding the boundary curve.

| $E_A$ | $\nu_A$ | $E_M$ | $\nu_M$ | $\varepsilon^L$ | $\left(\frac{\delta\sigma}{\delta T}\right)_L$ | $\sigma_L^S$ |
|---|---|---|---|---|---|---|
| 24 GPa | 0.33 | 35 GPa | 0.33 | 0.04 | $6.527 \frac{MPa}{°C}$ | 250 MPa |
| $\sigma_L^E$ | $T_0$ | $\left(\frac{\delta\sigma}{\delta T}\right)_U$ | $\sigma_U^S$ | $\sigma_U^E$ | $\sigma_{CL}^S$ | $\varepsilon_V^L$ |
| 270 MPa | 37 °C | $6.527 \frac{MPa}{°C}$ | 40 MPa | 20 MPa | 900 MPa | 0.04 |

**Table 1.** Superelastic nitinol stent material properties validated with *in vitro* data.

**RESULTS**

***Stent Geometry and Radial Force Validation***

Radial force was validated for the 23-mm and 26-mm CoreValve devices using *in vitro* data (Fig. 3b) [23]. The *in silico* radial force agrees well with the *in vitro* data for both stents, validating our nitinol material properties (Table 1). Manufactured by the same company, the remaining stent models were assumed to have the same material properties. Figure 3c shows the *in silico* curves of the radial force versus the stent crimping for all generations of the CoreValve and Evolut R devices. We see that larger models produce greater radial force at their respective deployment diameters. A sharp increase in radial force occurs across most of the stents within their respective deployment ranges.

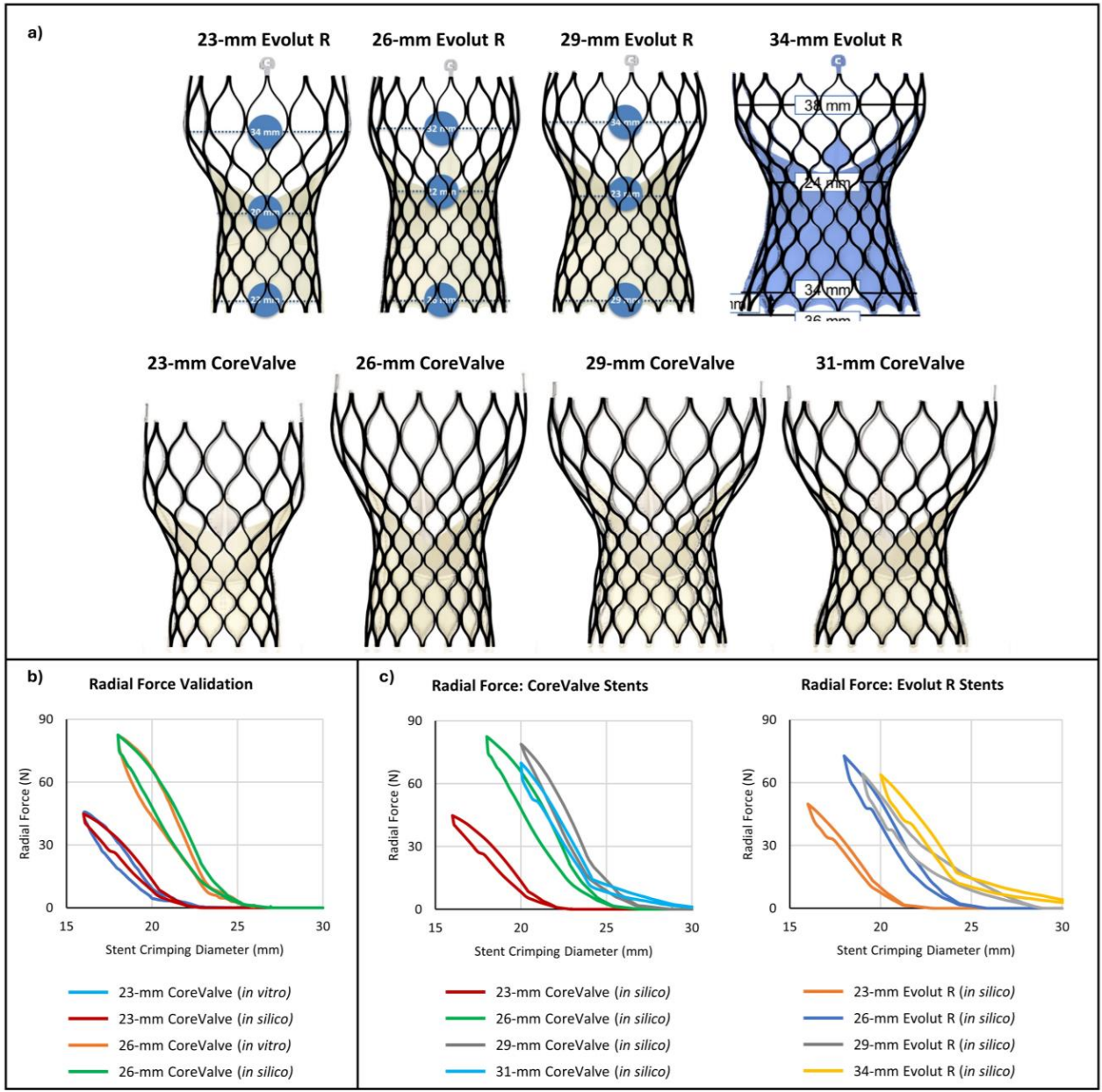

**Figure 3.** a) Evolut R (top row) and CoreValve (bottom row) geometry comparisons between *in silico* stent models (black, foreground) and the true stent models taken from rendered images (gray, background). b) Radial force validation results for both 23-mm and 26-mm CoreValve devices where the *in silico* radial force curves match the *in vitro* measurements very well. c) All *in silico* radial force measurements for CoreValve stent models (left) and Evolut R stent models (right).

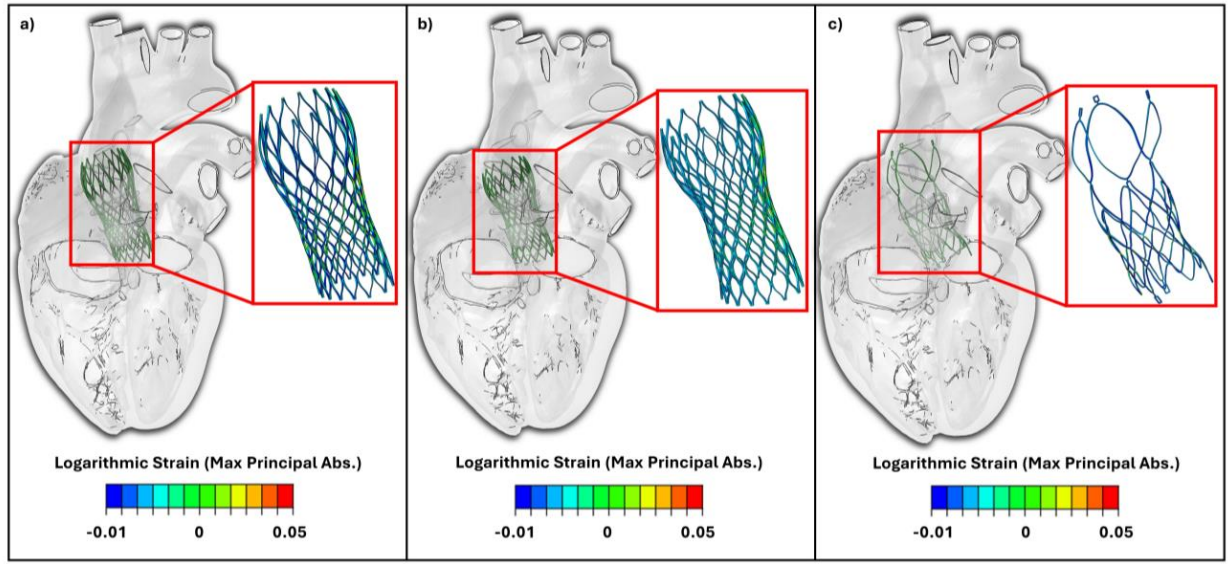

**Figure 4.** TAVR deployment images of stent models in the LHHM and associated snapshots of logarithmic max principal strains post-deployment for the a) 26-mm CoreValve, b) 26-mm Evolut and c) 27-mm PolyV-2 devices.

Once deployed, the stent frames experienced strain due to stent oversizing, anchoring within the native anatomy. Principal logarithmic strains ranged from -1% to 5% (Fig. 4). Tissue stresses and anchorage forces were analyzed to evaluate device deployment and effects on the surrounding myocardium. Figure 5 shows these results for the three control stent frames. The max principal stresses within the native tissue (Fig. 5, top row) was highest for the CoreValve, but all stent frames resulted in stresses below 2.5 MPa which has been established in the field as a threshold for tissue tearing [25]. Contact normal forces were observed in the lumen near or in contact with the stents (Fig. 5, bottom row). Performing a summation of these lumen nodal forces yielded total anchorage force magnitudes of 50 N, 40 N and 25 N for the CoreValve, Evolut R and PolyV-2, respectively. Contact normal forces were mainly exerted between the stent frames and the aortic annulus, native leaflets and aortic arch.

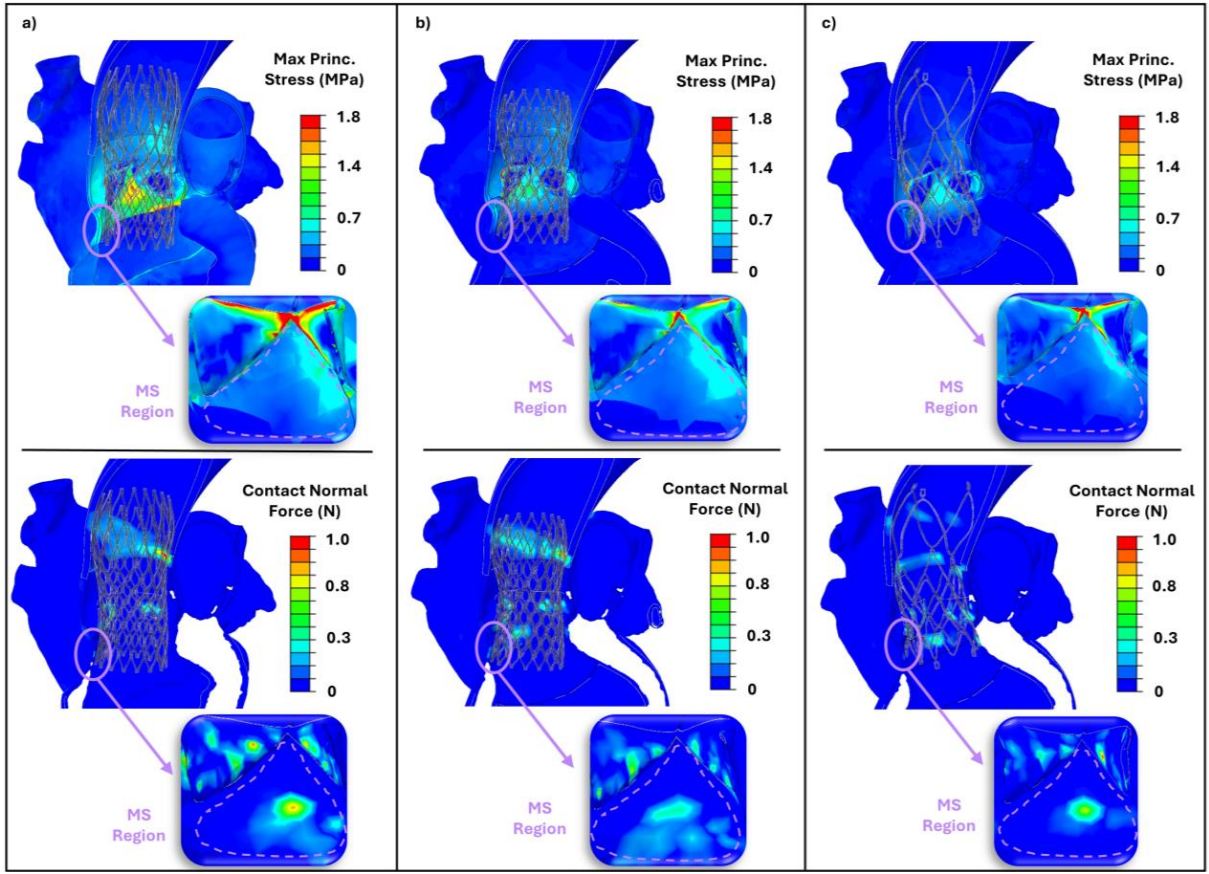

**Figure 5.** TAVR deployment images and contour plots for maximum principal stresses and contact normal forces in the lumen and membranous septum (MS) region for a) 26-mm CoreValve, b) 26-mm Evolut and c) 27-mm PolyV-2 devices.

| Stent Type | Control | 20% Increase | 20% Decrease |
|---|---|---|---|
| 26-mm CoreValve | CV-C | CV-20I | CV-20D |
| 26-mm Evolut R | EV-C | EV-20I | EV-20D |
| 27-mm PolyV-2 | PV2 | N/A | N/A |

**Table 2.** Stent model abbreviations used for describing fatigue analysis results. Each stent frame was named according to the stent model and width change of the struts.

*Stent Deformation Characteristics*

Device stent names and their abbreviations are summarized in Table 2 for clarity and conciseness. The motion of each was tracked, yielding compression, eccentricity, and radii deviation data over the third beat cycle (Supplementary Videos 4-6). Maximum contraction of the ventricles occurred at approximately 0.32 seconds during the cardiac cycle. At this timepoint, CV-C showed a compression, or change in average radius, of 3.5 ± 0.6 mm. However, CV-20I and CV-20D showed an average compression of 3 ± 0.6 mm and 4 ± 0.7 mm, respectively (Fig. 6).

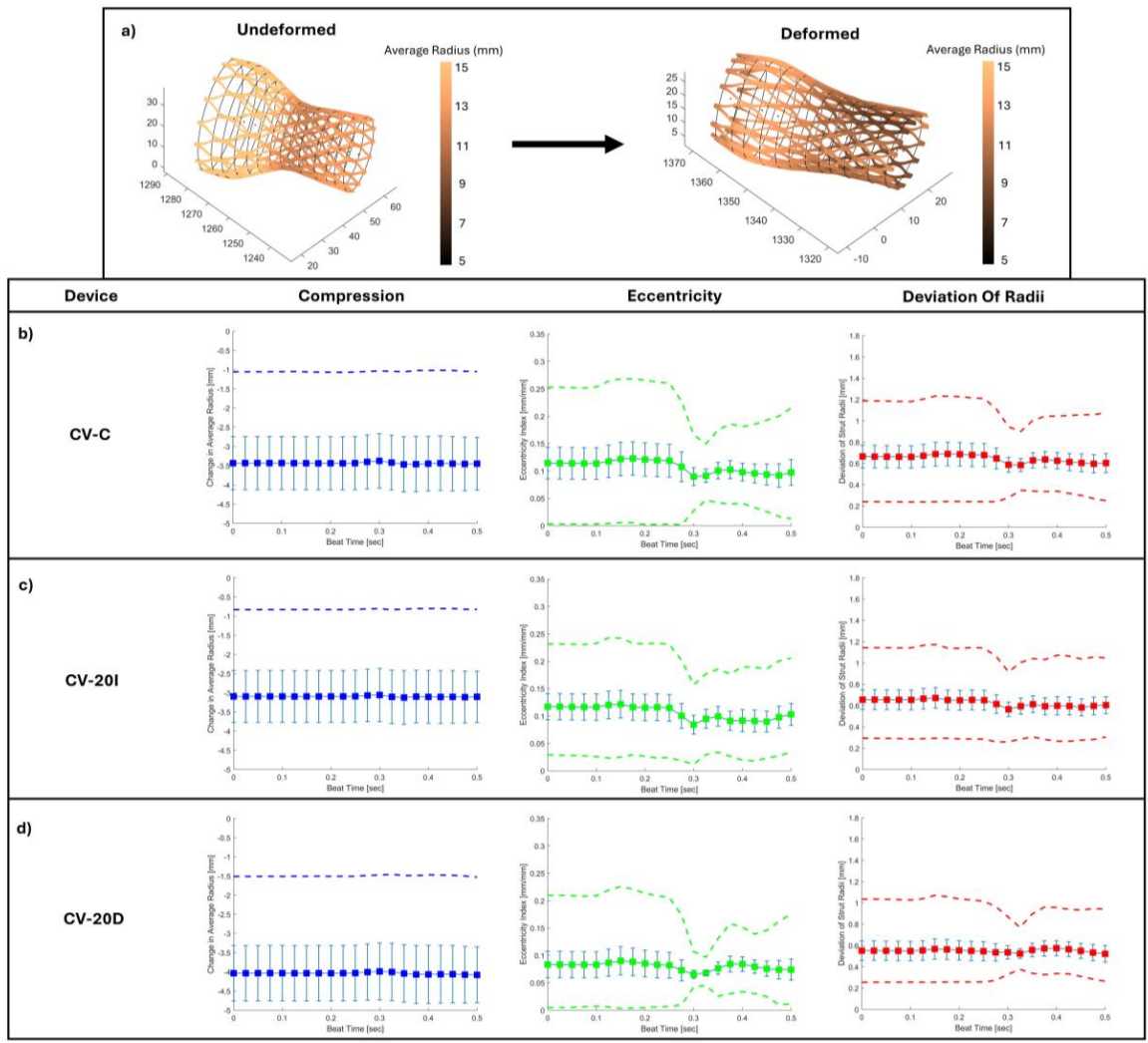

**Figure 6.** Stent tracking data for the CoreValve models. a) Undeformed and deformed images of the CV-C stent colored by average radius in mm. Plots for compression, or change in average radius, along with eccentricity index and deviation of strut radii are shown for b) CV-C, c) CV-20I and d) CV-20D.

When examining the eccentricity of the CoreValve stents, CV-C had an eccentricity index of 0.12 ± 0.03 at the beginning the third beat cycle, decreasing to 0.09 ± 0.01 during maximum contraction of the ventricles. CV-20I and CV-20D had eccentricity indices of approximately 0.12 ± 0.02 and 0.08 ± 0.01, respectively, at the start of the beat cycle, also decreasing to 0.10 ± 0.01 and 0.07 ± 0.01, respectively. The deviation of strut radii showed a similar trend for CV-C, in which the deviation started at 0.68 ± 0.1 mm and decreased to 0.58 ± 0.04 mm at peak contraction (Fig. 6). For CV-20I and CV-20D, the deviation of radii started as 0.65 ± 0.1 mm and 0.58 ± 0.08 mm, respectively. At peak contraction, these values decreased to 0.60 ± 0.04 mm and 0.57 ± 0.02 mm, respectively.

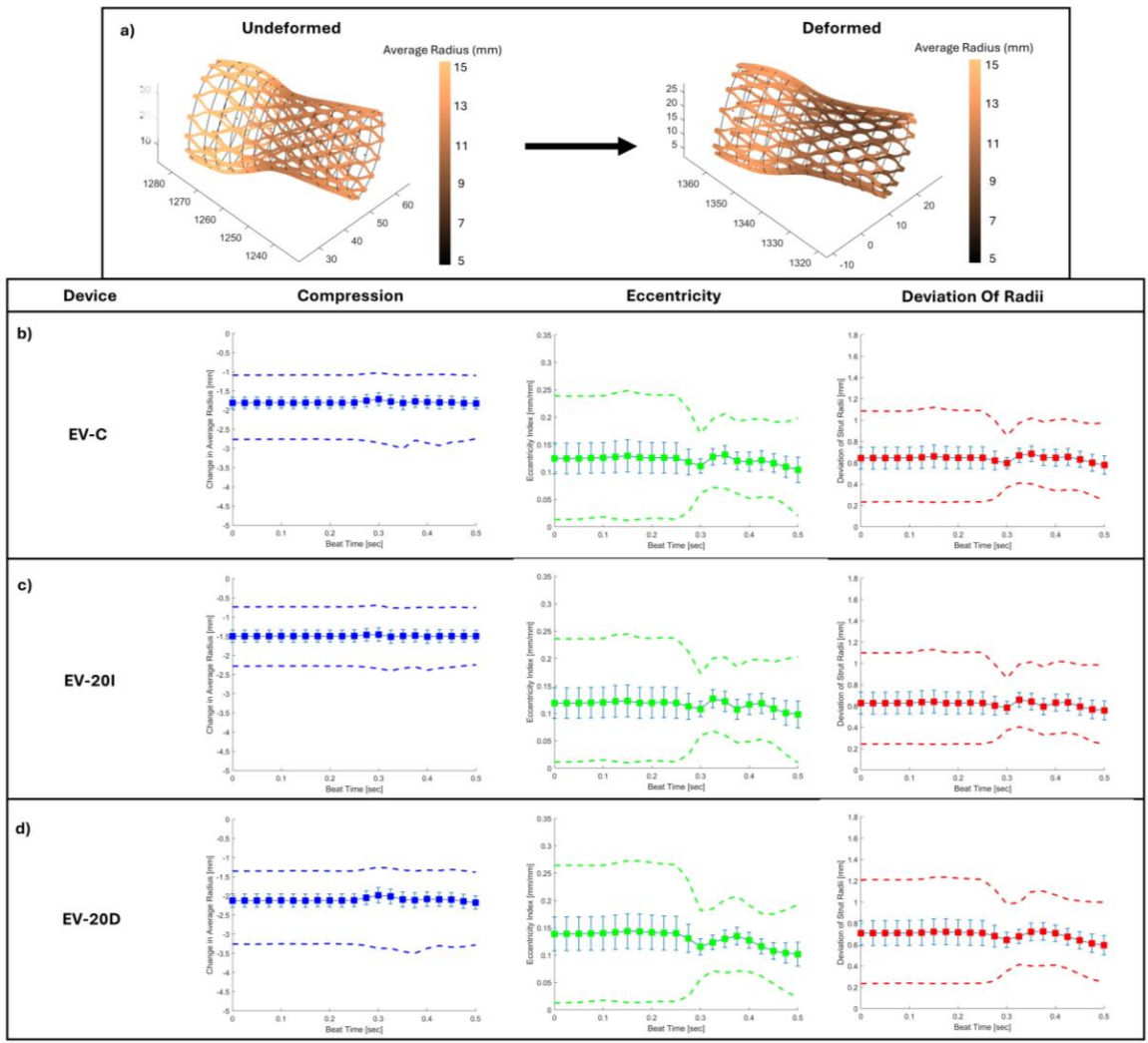

**Figure 7.** Stent tracking data for the Evolut R models. a) Undeformed and deformed images of the EV-C stent colored by average radius in mm. Plots for compression along with eccentricity index and deviation of strut radii are shown for b) EV-C, c) EV-20I and d) EV-20D.

EV-C was compressed, on average, 1.80 ± 0.10 mm during peak contraction of the heart (Fig. 7). However, EV-20I and EV-20D showed an average compression of 1.50 ± 0.10 mm and 2.10 ± 0.10 mm, respectively (Fig. 7). When examining the eccentricity of the Evolut R stents, EV-C had an eccentricity index of 0.13 ± 0.03 at the start of the beat cycle, oscillating about this value during peak contraction of the heart and eventually decreasing. EV-20I and EV-20D had eccentricity indices at the onset of the third beat of 0.12 ± 0.03 and 0.14 ± 0.04, respectively. During peak contraction, these values also oscillate, showing a steady decrease to 0.10 ± 0.03 and 0.11 ± 0.02, respectively. The deviation of strut radii for EV-C started at 0.64 ± 0.10 mm and maintained approximately the same value after max contraction of the heart. During the start of the beat cycle, EV-20I and EV-20D demonstrated deviation of radii at 0.61 ± 0.10 mm and 0.71 ± 0.10 mm, respectively. After some oscillation in these parameters during peak contraction, both values decreased slightly.

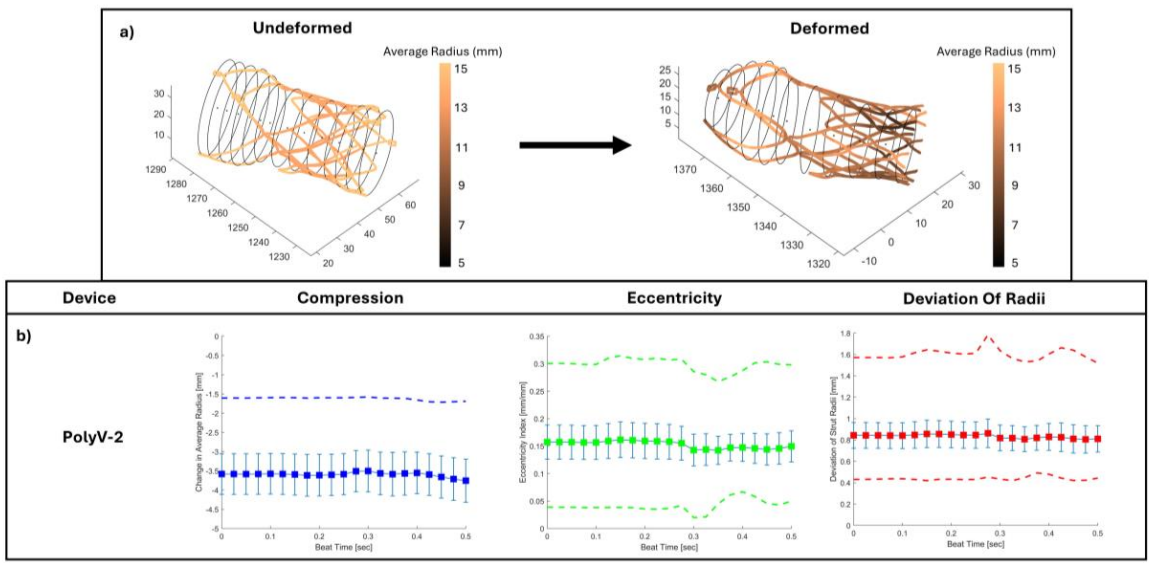

**Figure 8.** Stent tracking data for the PolyV-2 stent frame. a) Undeformed and deformed images of the PolyV-2 stent colored by average radius in mm. b) Plots for compression along with eccentricity index and deviation of strut radii are shown.

The PV2 data showed different behavior with a compression of 3.51 ± 0.50 mm (Fig. 8) and an eccentricity index of 0.16 ± 0.02 to begin the beat cycle, decreasing to about 0.14 ± 0.02 during peak contraction. The deviation of strut radii was maintained at 0.83 ± 0.10 mm.

*Fatigue Analysis of TAVR Stents*

The primary failure mode in this analysis was excessive strain amplitude (Supplementary Videos 7-9). Polar projections of strain amplitude for each stent model revealed the locations where element failure occurred, surpassing 0.4% (Fig. 9). There were common regions of high strain amplitude between all versions of the CoreValve and Evolut R devices, but upon close inspection, there were unique groups of elements that exhibit failure for each stent. PV2 experienced high strain amplitudes in localized regions as well but demonstrates less strain "hot spots" compared to the other devices. Each of the stent models were analyzed via sectioning (Fig. 10), as described below.

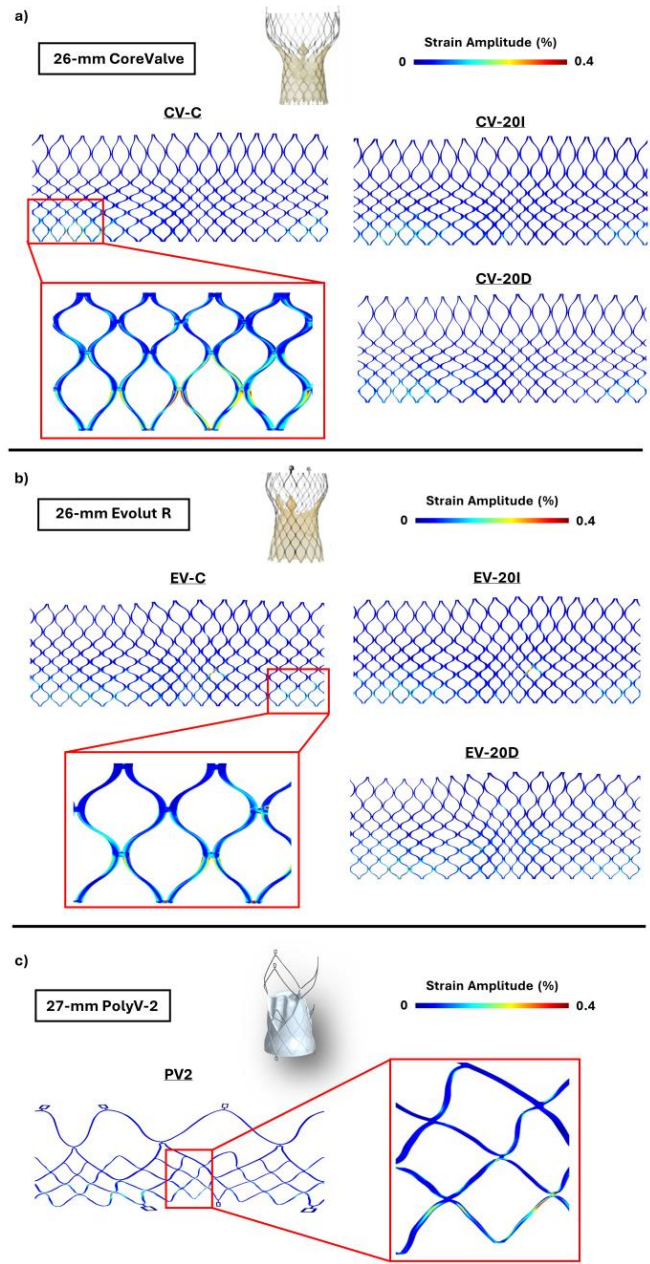

**Figure 9.** Polar projections of strain amplitude. There are common regions of high strain amplitude between all versions of the CoreValve and Evolut R devices, but due to differences in strut width and stent design, there are unique groups of elements that exhibit failure for each stent frame. The PolyV-2 stent frame experienced high strain amplitude in localized regions as well but demonstrates superior fatigue resistance comparatively.

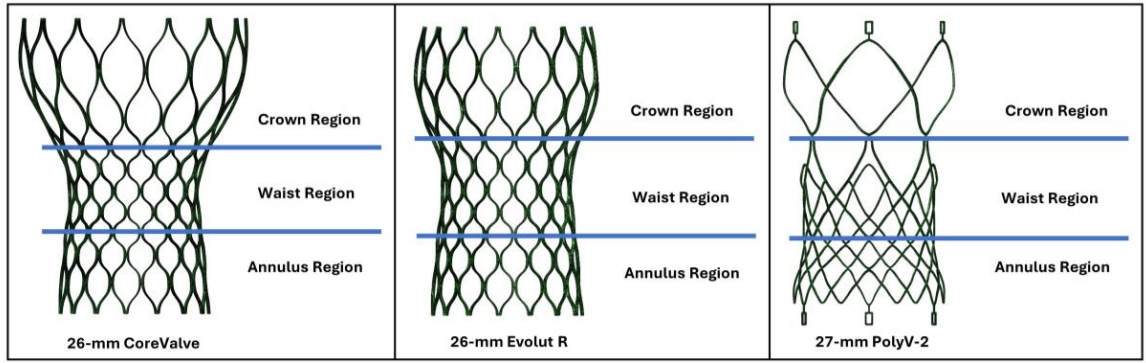

**Figure 10.** Defined stent regions in which mean strain and strain amplitude were analyzed post-TAVR.

*Crown Region Fatigue*

Among all variations of the 26-mm CoreValve, no crown elements showed excessive strain amplitudes or mean strains beyond the failure threshold (Fig. 11). The crown elements for CV-C displayed the highest strain amplitude, reaching about 0.3% while the crown elements in CV-20I demonstrated the highest mean strain at 5%. The crown elements from CV-20D never exceeded mean strains of more than 4%. The 26-mm Evolut R models displayed a different fatigue distribution, but these elements also showed no indication of failure. EV-20D showed the highest strain amplitude at about 0.38%, while the max mean strain reached approximately 4.3% in EV-20I. The crown elements of PV2 showed superior fatigue resistance. No elements surpassed the failure curve, barely reaching 0.2% and 2% for strain amplitude and mean strain, respectively (Fig. 11), as described below.

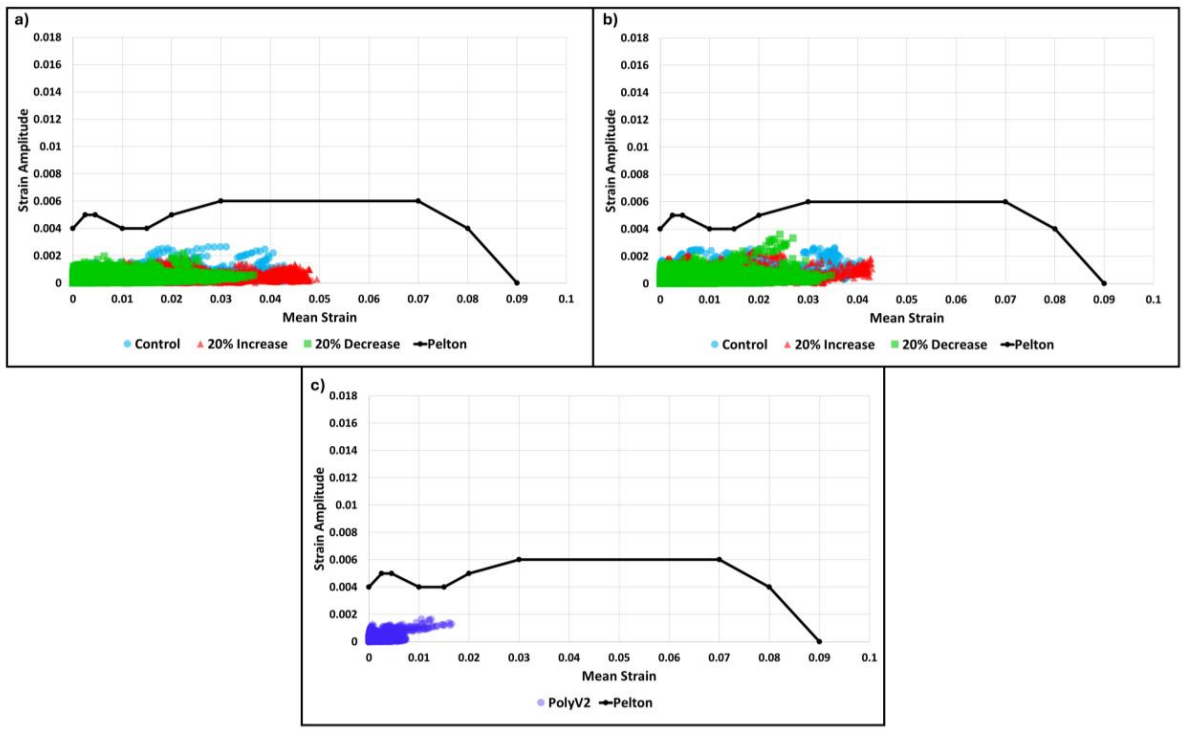

**Figure 11.** Strain data for the fatigue analysis of the crown regions for the a) CoreValve stent models b) Evolut R stent models and c) PolyV-2 stent model.

*Waist Region Fatigue*

The waist elements showed much higher magnitudes of strain amplitude and mean strain (Fig. 12). CV-20I and CV-20D displayed failing elements due to excessive strain amplitudes in various elements. For CV-20D, approximately 15 elements showed strain amplitudes above 0.8% with mean strains ranging from 2% to nearly 5%. At lower mean strains, CV-C also showed element failure due to many elements displaying strain amplitudes over 0.4%. However, CV-20I showed no indication of element failure with strain amplitudes well below 0.4% and mean strains reaching 4.5%. EV-C and EV-20I showed similar results in terms of maximum strain amplitudes and mean strains as they both attained strain amplitudes of about 1.6% at mean strains up to 5%. EV-20D also showed element failure with strain amplitudes reaching 0.8% at low mean strains. Unlike the previous stent models, the waist elements of PV2 did not show any indication of fatigue failure. The PV2 waist elements showed low strain amplitudes and mean strains, barely reaching 0.4% and 6.5%, respectively.

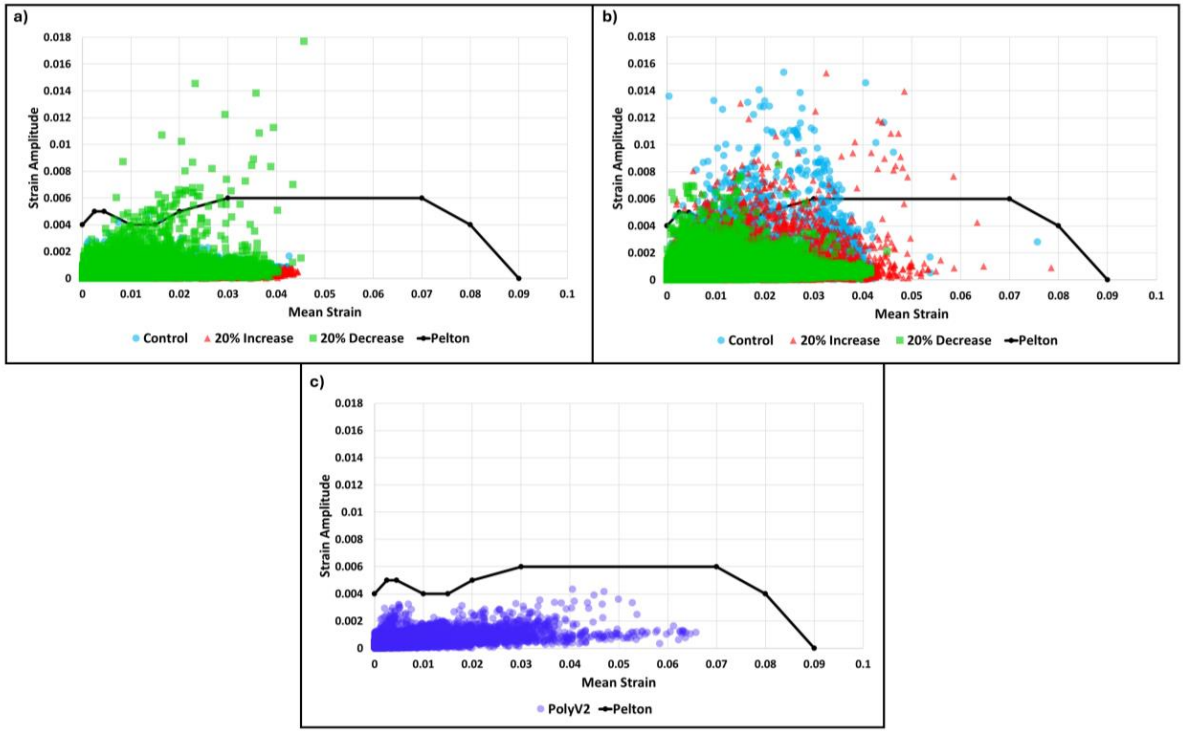

**Figure 12.** Strain data for the fatigue analysis of the waist regions for the a) CoreValve stent models b) Evolut R stent models and c) PolyV-2 stent model.

*Annulus Region Fatigue*

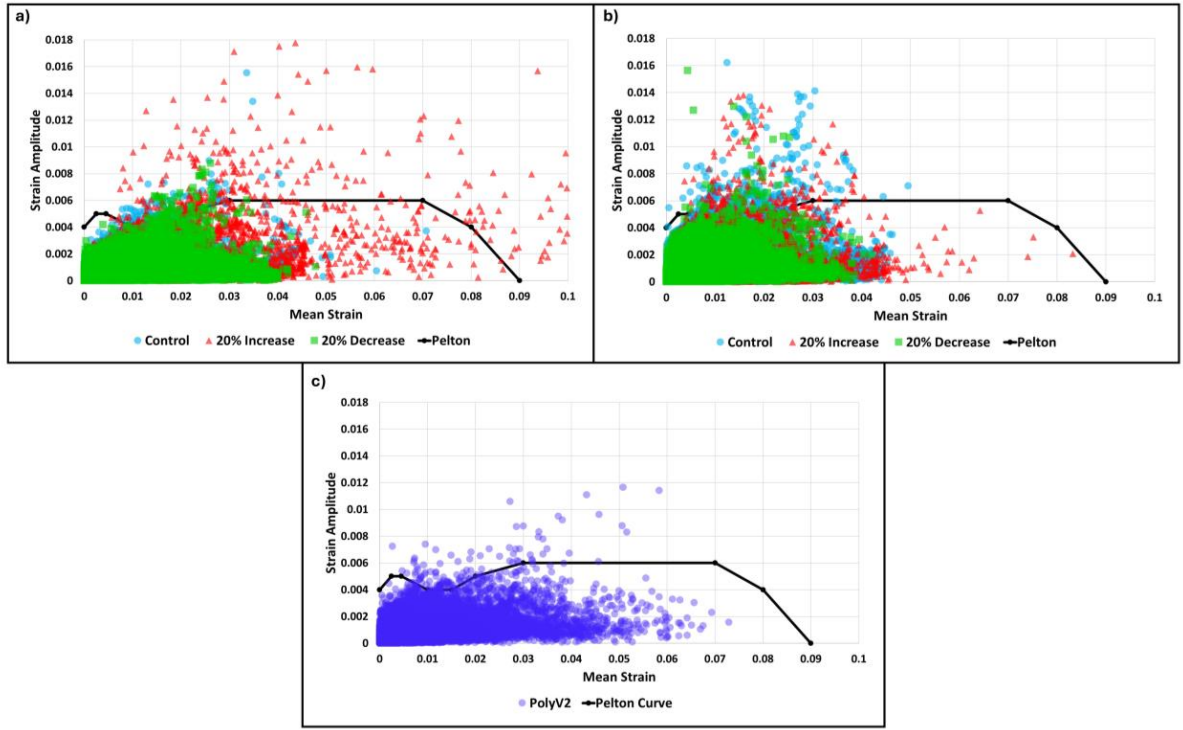

**Figure 13.** Strain data for the fatigue analysis of the annulus regions for the a) CoreValve stent models b) Evolut R stent models and c) PolyV-2 stent model.

Hundreds of CV-20I annulus elements surpassed the fatigue failure threshold, where most of these elements exhibited strain amplitudes ranging from 0.6% to 1.2%. Nearly 30 elements failed due to excessive mean strains in the range of 7% to 10%. CV-20D and CV-C did not show this magnitude of mean strain but did display high strain amplitudes, ranging from 0.4% to 0.8%. Although very similar, CV-20D in general, had smaller strains than CV-C. All three Evolut R stent models showed a high degree of element failure when analyzing the annulus elements. EV-C, EV-20D and EV-20I all showed similar results in terms of maximum strain amplitudes, with most of the data beyond the fatigue threshold in the 0.6% to 1.4% range. The maximal mean strain for EV-C and EV-20D did not exceed 5%, while EV-20I had nearly 20 elements beyond 5% mean strain. The annulus elements of PV2, unlike the fatigue data in the previous two regions, did show indications of failure due to excessive strain amplitudes reaching 1.2% at mean strains ranging from 1% to 6%.

**DISCUSSION**

TAVR stent manufacturers do not publicly disclose their device material properties. This mandates that the material parameters used in the models should be independently extracted. Accordingly, validation study of the stents was conducted to ensure that the virtual models accurately represent the commercial devices [23]. Our nitinol material parameters differ from other material properties specified in the literature for computational TAVR studies using nitinol stents [25, 26]. Other methods for determining superelastic material properties have been explored such as pairing constitutive models with experimental testing of thin nitinol wire [27, 28]. The variation in material properties reported in TAVR research is expected, due to the unique manufacturing processes and temperature treatments used [23]. Our validation study was tailored to the specific devices that were reconstructed using radial force data instead of using parameters based off nitinol wire that would likely lead to unrealistic TAVR device deployment and fatigue data.

The stent deformation and tracking algorithms used here are novel components of this workflow that allows for crucial geometric information to be acquired after deployment. Each stent was compressed differently upon deployment and demonstrated varying degrees of eccentricity and variation of radii. As the stents experience complex, non-uniform cardiac loading, these parameters fluctuate, yielding valuable information about stent structural response during peak contraction and recovery of the heart that may be used for engineering optimization processes.

Previous fatigue analyses and testing protocols fail to capture accurate loading conditions. The current analysis offers a unique tool to analyze boundary conditions that closely mimic *in vivo* loading that in turn may obviate costly human or animal trials. It establishes an *in silico* platform to study the impact of device design features on device performance, and a more realistic study of the interaction of TAVR devices with cardiac structure and motion. Previous studies examining arterial stent fatigue via simplified approaches that computed maximum mean strains and strain amplitudes of 1.22% and 0.24%, respectively, under so-called physiological pressure conditions [7]. However, these values are substantially lower than the data recorded in our study. Excessive strain amplitudes were the primary cause of element failure in this analysis, indicating a large variation in strain experienced by TAVR stents. We note that the crown region had minimal strains in all stent models, but upon analyzing the waist and annulus regions, we see that all stent models had strain values exceeding conventional fatigue limits for nitinol. These observations from the data are consistent with more recent analysis of fatigue strength and reliability of arterial stents [29].

Our results highlight how crucial geometric parameters are for evaluating actual device performance. We analyzed how stent compression, eccentricity, and deviation of radii all change when stent design is altered. Such parameters ultimately determine the strains experienced by stents after TAVR deployment and change in the strain distribution. When comparing the older generation CoreValve design to that of the newer generation Evolut R, we may at first suppose that the Evolut R offers superior performance due to stent tracking results. However, our study approach clearly indicates that the Evolut R demonstrated the greatest number of failing elements in all regions. This was observed due to the higher strain amplitudes experienced by the Evolut R device resulting from the smaller device profile and relative under-expansion, leading to varying contact with the lumen throughout the cardiac cycle. These observations agree with results published in a similar, previous study [30], additionally highlighting the importance of capturing the dynamic interaction between the stent and the aortic root. The results also highlight that strut width, along with other modifiable parameters of a stent design, may not be able to directly predict *in vivo* strain behavior [7, 29]. This suggests that the strain experienced by TAVR stents is extremely complex and may not be expressed in terms of a single equation or relation, but rather, must be assessed by studying the stent response under physiological cardiac loading. This demonstrates the utility of advanced *in silico* analysis of TAVR devices with the LHHM and supports the future of era of clearance and approval by regulatory bodies such as the Food and Drug Administration (FDA) using *in silico* testing for medical devices [23].

Our study had several limitations. The LHHM represents a healthy adult human heart, and to study TAVR device performance more accurately, a diseased condition should be introduced, such as adding calcifications on the aortic leaflets [31]. Since our study is comparative, such a limitation does not hinder the validity of the results, but rather highlights the need for models used to analyze TAVR stents fatigue under loading conditions closer to the clinical scenario. Currently, our modeling approach cannot predict the time until failure, but such formulations that consider cumulative effects that may lead to stent failure can be included in future work. For the specific case of the PolyV-2 stent frame, due to the novelty of this device, the data needed for radial force validation study is still pending. Currently this was mitigated by using the material properties validated for the CoreValve and Evolut R devices for the PolyV-2. Lastly, the limited fatigue life of the TAVR stents indicated by our rigorous analysis highlights a clinical complication that may eventually develop as younger, lower-risk TAVR patients, age. Given the scarcity of fatigue data currently available in the field, our study highlights the urgent need for more computational fatigue studies.

**CONCLUSION**

We presented a first-of-its-kind analysis of TAVR stent deployment under close to realistic cardiac loading conditions during multiple cardiac cycles, using the SIMULIA Living Heart Human Model. This highlights the utility of such *in silico* tool that indicate specific stent regions that could become prone to failure, by allowing an unparalleled design optimization approach for TAVR devices, such as our polymeric TAVR device. It offers a unique platform to estimate TAVR stent fatigue that could supplement a multitude of other device performance parameters to include potential failure modes. This complementary approach to *in vitro* durability testing may eventually be established as a cost-effective alternative to it. It allows for a more realistic study of the dynamic interaction of TAVR devices with cardiac structures and tissues. Our study indicates specific modes of potential stent failure in current commercial devices, and optimization approaches to mitigate them. Future studies will incorporate novel, patient-specific beating heart models with morphable geometry capabilities, including calcifications.

**ACKNOWLEDGEMENTS**

Funding: This work was supported by the National Institutes of Health grants: NIBIB: BRP U01 EB026414 (DB), and NHLBI: R41 (DB) HL134418 and R42 HL134418 (DB).

Industry Partners: ANSYS, SIMULIA Living Heart Project

Disclosure: Author BK is a consultant for PolyNova Cardiovascular Inc. Authors DB and MS have an equity interest in PolyNova Cardiovascular Inc.